\documentclass[english,english,noshowpacs,twocolumn,nofootinbib,amsmath,amssymb,aps,pra]{revtex4-2}

\usepackage{amssymb}
\usepackage{amsthm}
\usepackage{amsfonts}
\usepackage{amsmath}
\usepackage{braket}
\usepackage{natbib}
\usepackage{graphicx}
\usepackage{xcolor}
\usepackage{soul}

\begin{document}

\title{Effect of Closely-Spaced Excited States on Electromagnetically Induced Transparency}
\author{Saesun Kim}%
 \email{saesun.kim-1@ou.edu}
 \author{ Alberto M. Marino}%
 \email{marino@ou.edu}
\affiliation{Homer L. Dodge Department of Physics and Astronomy, University of Oklahoma, Norman, OK 73019\\
Center for Quantum Research and Technology, University of Oklahoma, Norman, OK 73019}

\begin{abstract}

Electromagnetically induced transparency (EIT) is a well-known phenomenon due in part to its applicability to quantum devices such as quantum memories and quantum gates. EIT is commonly modeled with a three-level lambda system; however, this simplified model does not capture all the physics of EIT experiments with real atoms. We present a theoretical study of the effect of two closely-spaced excited states on EIT and off-resonance Raman transitions. We find that the coherent interaction of the fields with  two excited states whose separation is smaller than their Doppler broadened linewidth can enhance the EIT transmission and broaden the width of the EIT peak.  However, a shift of the two-photon resonance frequency for systems with transitions of unequal dipole moments leads to a reduction of the maximum transparency that can be achieved when Doppler broadening is taken into account even under ideal conditions of no decoherence.  Only when the separation between the two excited states is of the order of the Doppler width or when the transitions have the same dipole moments can complete transparency be recovered. We present experimental EIT measurements in the D1 lines of $^{85}$Rb and $^{87}$Rb that agree with the theoretical predictions and present a dress state picture to explain the observed behavior in all the transitions of the D1 lines of both Rb isotopes that result from the two closely-spaced excited state. Finally, we show that off-resonance Raman absorption is enhanced and its resonance frequency is shifted.

\end{abstract}

\maketitle

\section{\label{Intro}Introduction}

Electromagnetically induced transparency (EIT) and off-resonant Raman transitions are established techniques to implement  optical quantum memories~\cite{EITHarris1,EITHau,EITPeng,OFFGUO,OFFWalmsley,OFFWalmsley2} and quantum gates~\cite{GateLukin,GateDurr,OffMonroe}. Commonly, these processes are modeled with a three-level lambda configuration~\cite{EITXiao1,EITXiao2,EITWalsworth,ARIMONDO1996257}; however, all of the D1 transitions in alkali atoms have four hyperfine levels. As a result, it is necessary to consider the effect of the two excited states whose frequency separation is smaller than or of the order of the Doppler broadening when working with atomic vapors.

Several papers have previously studied coherent atom-photon interactions in multilevel atomic systems. In these systems, the cancellation of spontaneous emission due to interference of the two excited states has been predicted and demonstrated~\cite{SponScully,SponZhu}.  Additionally, shifts of the two-photon resonance in EIT and Raman absorption in a multilevel system have been theoretically predicted and observed experimentally~\cite{SlowBehroozi,SlowHam}. A significant reduction of the EIT transmission in a system with four closely-spaced excited states has been theoretically predicted~\cite{EITGiacobino1}, in agreement with EIT experiments in the D2 line of alkali atoms~\cite{EITGiacobino2,EITPayne,EITLezama,EITGhosh}. These studies indicate that the physics of EIT in real atomic systems is much richer than the one predicted with a simple three-level model.

In this paper, we focus on the theoretical study of EIT and off-resonant Raman transitions in a system with  two closely-spaced excited states through a model based on a four-level system. We derive analytical expressions for the atomic susceptibilities with Doppler broadening using the density matrix formalism under the assumption of a weak probe field. In order to identify effects due to the two closely-spaced excited levels, we compare the predictions from the four-level system with  ones based on a model composed of two independent three-level systems, as shown in Fig.~\ref{Figure1}.

From this comparison, we find several interesting results that are due to having two excited levels that coherently interact with the optical fields. In the limit of a large decoherence rate, which corresponds to the limit in which superconducting circuits~\cite{PhysRevLett.93.087003}, artificial meta-atoms~\cite{article3}, and quantum dot~\cite{article5} are modeled, the interaction between the fields and the two excited states leads to an enhancement of the EIT transmission. However, an unequal dipole moment of the transitions between the ground states and the excited states leads to a shift of the two-photon resonance that makes it impossible to obtain perfect transparency when Doppler broadening is taken into account, even in the ideal case of no decoherence, which is consistent with Ref~\cite{EITGiacobino1}.

We further investigate the underlying physical process through a dressed state model and find that this reduction does not occur when the separation between the excited states is larger than the Doppler broadening or when the dipole moments of the transitions are equal. We then use this model to study the effect of the spacing between the two excited states, decoherence rate, and Rabi frequency on the EIT transmission. In addition, we show that the presence of two closely-spaced excited states leads to an enhancement of the off-resonance Raman absorption and a shift of its resonance frequency. Finally, we present experiments done in the D1 lines of $^{85}$Rb and $^{87}$Rb and use our dressed state model to explain how the two closely-spaced excited state leads to the observed behaviors in the different transitions of the D1 lines.

\section{\label{TheorM}Theoretical Model}

\begin{figure}[htb]
  \centering
  \includegraphics{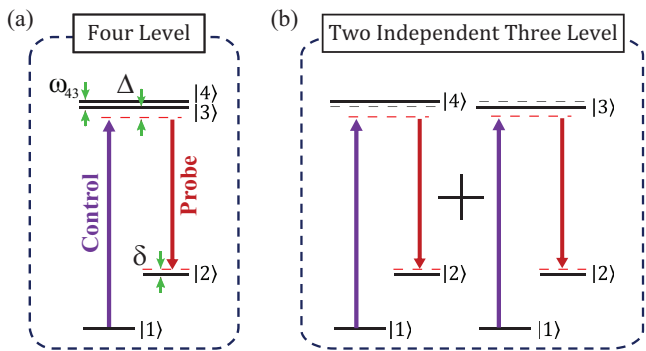}
  \caption{Schematic diagrams of the energy level structures used to model EIT in the D1 line of an alkali atom. We describe the system with (a) a four-level model and (b) a model composed of two independent three-level systems. Energy levels $\ket{1}$ and $\ket{2}$ are the two ground hyperfine states, while $\ket{3}$ and $\ket{4}$ are the two excited hyperfine states. The one-photon detuning and two-photon detuning are labeled as $\Delta$ and $\delta$, respectively.}
  \label{Figure1}
\end{figure}

To model EIT and Raman transitions in the D1 line of alkali atoms we consider the four-level model shown in Fig.~\ref{Figure1}(a). A control field couples level $\ket{1}$ with the two excited states $\ket{3}$ and $\ket{4}$, while a probe field couples level $\ket{2}$ with the two excited states. The transitions between the two lower levels $\ket{1}$ and $\ket{2}$ and between the two upper levels $\ket{3}$ and $\ket{4}$ are taken to be dipole forbidden. We define the one-photon detuning for the control field as $\Delta=\omega_{31}-\omega_{c}$ and the two-photon detuning as $\delta=\omega_{21}-\omega_{p}+\omega_{c}$, where $\omega_{ij}$ is the transition frequency between energy levels $\ket{i}$ and $\ket{j}$ and $\omega_{c}$ and $\omega_{p}$ are the frequencies of the control and probe fields, respectively. The Rabi frequencies of the control and probe fields are defined as $\Omega_{ij} \equiv 2 d_{ij} E/\hbar$, where $d_{ij}$ is the dipole moment between levels $\ket{i}$ and $\ket{j}$, and $E$ is the amplitude of the electric field that couples that transition. In order to identify the effects due to the coherent coupling of the two ground states with the two closely-spaced excited states through the control and probe fields, we introduce a model composed of two independent three-level systems, one with its excited state at the same energy as level $\ket{4}$ and the other with its excited state at the same energy as level $\ket{3}$, as shown in Fig.~\ref{Figure1}(b). We then combine the response of these two independent three-level systems to obtain an effective response for the four levels that does not allow for an effective interaction between the two excited states.

We use the density matrix formalism with spontaneous emission and collisional damping~\cite{QuanOptKnight,AtomphotonGrynberg} to calculate the equations of motion. In the rotating wave approximation and weak probe field limit, we calculate the steady-state solution to obtain the atomic coherence for each transition, from where we obtain analytical expressions for the susceptibilities $\chi_{ij}$ for the two models. The steady-state solutions are then used to obtain analytical solutions that take into account Doppler broadening by integrating over the Maxwellian velocity distribution following the procedure outlined in~\cite{EITXiao1}. Detailed calculations and analytical solutions are given in Appendices~\ref{app:A} and~\ref{app:B}. By comparing the response of the four-level and the two independent three-level models, we can gain an understanding of the effect of having fields that can coherently interact with two closely-spaced excited states. While the results are valid for any four-level system, such as the D1 line of alkali atoms, in the following subsections we specialize to the case of $^{85}$Rb for which the natural linewidth $\gamma$ of the D1 line is $5.75$~MHz and the separation between the two excited states is $361$~MHz.

\subsection{\label{TheorM:A}Electromagnetically induced transparency}

\begin{figure}[htb]
  \centering
  \includegraphics{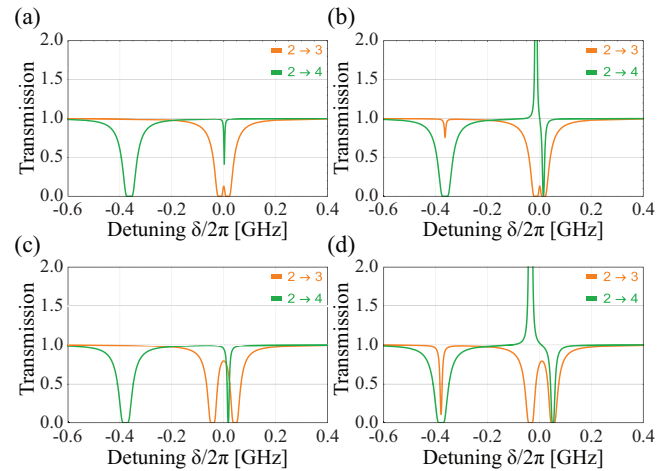}
  \caption{Theoretical transmission spectra for the probe field obtained from the calculated susceptibilities $\chi_{32}$ and $\chi_{42}$ (see Appendix~\ref{app:A} for more detail) as a function of the two-photon detuning $\delta$/2$\pi$. We consider the response for the two independent three-level model for (a) $\Omega_{31}=5\gamma$ and (c) $\Omega_{31}=15\gamma$, and for the four-level model for (b) $\Omega_{31}=5\gamma$ and (d) $\Omega_{31}=15\gamma$. For all the transmission spectra we take $T=50^{\circ}$C, $\Delta=0$, $\gamma_{21}=0.25\gamma$ (see Sect.~\ref{ExperiR:A}), and a cell length of 2.54~cm.}
  \label{Figure2}
\end{figure}

We first study the effect of the two closely-spaced excited levels on EIT. Figure~\ref{Figure2} compares the probe transmissions obtained from the calculated susceptibilities for the four-level model and the two independent three-level model once propagation through the atomic medium is taken into account. Each figure contains two transmission traces, one for the $\ket{2}$ to $\ket{4}$ transition, calculated from the susceptibility $\chi_{42}$ (green line), and the other for the $\ket{2}$ to $\ket{3}$ transition, calculated from the susceptibility $\chi_{32}$ (orange line).

We start by considering the response of the two independent three-level model, shown in Fig.~\ref{Figure2}(a). In this case, EIT is a result of the response due to $\chi_{32}$ (orange line) when the fields are tuned to two-photon resonance ($\delta=0$). On the other hand, the response due to $\chi_{42}$ (green line) leads to Raman absorption at two-photon resonance. The combination of Raman absorption and EIT at the same frequency leads to a reduction of the overall transmission once the full response of the atom is considered. When the Rabi frequency of the control field is significantly larger than the decay rate $\gamma$, Autler-Townes splitting centered at $\delta=0$ appears for the response due to $\chi_{32}$ (orange line), see Fig.~\ref{Figure2}(c).  However, the off-resonance Raman absorption due to the response from $\chi_{42}$ (green line) experiences a light shift that moves the resonance to the red of $\delta=0$. As a result, Raman absorption and EIT now occur at different two-photon detunings, which minimizes the cancellation in transparency present at smaller Rabi frequencies.

We next consider the response of the four-level model in Fig.~\ref{Figure2}(b). Similar to the two three-level model, EIT appears on the $\ket{2}$ to $\ket{3}$ transition when the control and probe fields are on two-photon resonance, as can be seen from the response due to $\chi_{32}$ (orange line) in Fig.~\ref{Figure2}(b). However, the $\ket{2}$ to $\ket{4}$ transition now exhibits gain to the blue and  Raman absorption to the red of the two-photon resonance, as shown by the response due to $\chi_{42}$ (green line) in Fig.~\ref{Figure2}(b). Since now EIT and Raman absorption appear at different two-photon detunings and there is gain close to the two-photon resonance, the EIT transmission is enhanced. For the four-level model there is an extra absorption feature from the response due to $\chi_{32}$ (orange line) that appears when the probe is on resonance with level $\ket{4}$ ($\delta=-361$MHz), which leads to an increased absorption of the probe when it is on resonance with the $\ket{2}$ to $\ket{4}$ transition. For larger control field Rabi frequencies, the EIT peak is red shifted with respect to two-photon resonance ~\cite{SlowBehroozi,SlowHam}, as can be seen by the response of $\chi_{32}$ (orange line) in Fig.~\ref{Figure2}(d). However, the gain and Raman absorption in the response due to $\chi_{42}$ (green line) shift by different amounts around the two-photon resonance. As a result, the gain partially cancels the absorption to the blue of the two-photon resonance and the Raman absorption partially enhances the absorption to the red of the two-photon resonance. The combination of these two effects leads to an asymmetric EIT lineshape and a broadening of the EIT peak.

\begin{figure}[htb]
  \centering
  \includegraphics{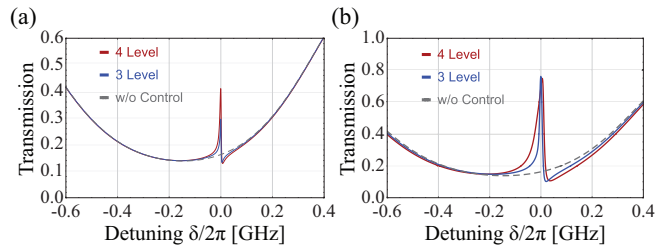}
  \caption{Doppler broadened EIT transmission spectra as a function of the two-photon detuning $\delta$/2$\pi$ for (a) $\Omega_{31}=5\gamma$ and (b) $\Omega_{31}=15\gamma$.  The full response of the atomic system is obtained by adding the susceptibilities of two transitions the probe can couple to, $\chi_{42}$ and $\chi_{32}$ (see more detail in Appendix~\ref{app:B}). The response is calculated for $T=50^{\circ}$C, $\Delta=0$, $\gamma_{21}=0.25\gamma$ (see Sect.~\ref{ExperiR:A}), and a cell length of 2.54~cm.}
  \label{Figure3}
\end{figure}
We finally consider the effect of Doppler broadening for both the four-level and the two three-level models, see Appendix~\ref{app:B} for the derivation of analytical expressions.  The responses for the two models after propagation through the atomic medium taking into account the calculated susceptibilities are shown in Fig.~\ref{Figure3}, where the transmission spectra for the four-level model (red line), for the two three-level model (blue line), and in the absence of the control field (dashed line) are plotted for a temperature ($T$) of $50^{\circ}$C, which corresponds to a number density $\approx$ 1.5*$10^{17}$m$^{-3}$. Since the separation between the excited states of $^{85}$Rb is 361 MHz and the Doppler width at 50$^{\circ}$C is 527 MHz, the two excited hyperfine levels ($\ket{3}$ and $\ket{4}$) are not resolved. For the case of low Rabi frequency for the control field, see Fig.~\ref{Figure3}(a), EIT has a higher transmission for the four-level model than for the two independent three-level model. This is due to the fact that Raman absorption for the four-level system is shifted significantly further from the two-photon resonance than for the two three-level model. Additionally, the presence of gain near two-photon resonance due to the simultaneous coupling of the fields to the two excited levels in the four-level model enhances the EIT transmission.

For large Rabi frequencies of the control field, the Raman absorption for the two three-level model shifts enough from the two-photon resonance that it no longer reduces the EIT transmission. On the other hand,  for the four-level model Raman absorption and EIT  shift together. As a result, the EIT transmission is always limited by Raman absorption in spite of the presence of gain. In this regime, the gain to the blue of the EIT peak starts to broaden, which leads to an effective broadening of the transmission peak and an asymmetric lineshape, as can be seen in Fig.~\ref{Figure3}(b). We find that the EIT peak from the four-level model is broadened more than the one from the three-level model due to the enhanced Raman absorption to the red of the two-photon resonance from the response due to $\chi_{42}$ and the additional absorption for the $\ket{2}$ to $\ket{4}$ transition due to the response from $\chi_{32}$.

\subsection{\label{TheorM:B}Role of Rabi frequency and decoherence}

We next compare the dependency of the maximum EIT transmission and width of the transparency peak on the control field Rabi frequency and decoherence rate for the four-level and the two three-level models.  This comparison allows us to study the effect of the coherent coupling of the fields to the two excited states in the four-level model and its impact on  EIT. To do this, we define the transparency as $1-\textrm{Im}[\chi_\textrm{EIT}]/\textrm{Im}[\chi_\textrm{Abs}]$ where $\chi_\textrm{EIT}$ and $\chi_\textrm{Abs}$ are the Doppler-broadened susceptibilities at the frequency of the maximum EIT transmission with and without the control field, respectively. The transparency as defined here provides a measure of the EIT contrast based on the absorption coefficients.  We define the EIT width as the FWHM of the EIT transmission peak.

\begin{figure}[htb]
  \centering
  \includegraphics{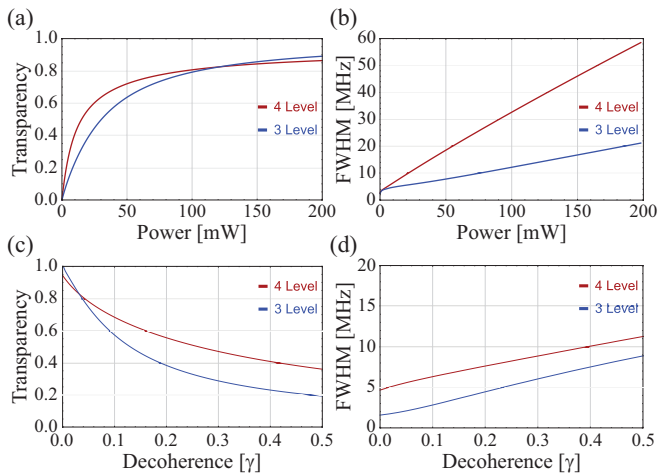}
  \caption{Dependence of  EIT on the Rabi frequency of the control field and decoherence rate between the two ground states for the Doppler broadened four-level and two three-level models. (a) EIT transparency and (b) FWHM of the EIT transmission peak as a function of control field power for a decoherence rate of $\gamma_{21}=0.25\gamma$ (see Sect.~\ref{ExperiR:A}). (c) EIT transparency and (d) FWHM of the EIT transmission peak as a function of the decoherence rate between the two ground states for a control field Rabi frequency of $\Omega_{31}=5\gamma$. For all the figures, the calculations are done for $T=50^{\circ}$C, $\Delta=0$, a diameter of the control field of 1~mm (which results in a Rabi frequency of $\sim 18\gamma$ for a power of 200~mW), and a cell length of 2.54~cm.}
  \label{figure5}
\end{figure}
We first compare the transparency for the two models. As can be seen in Fig.~\ref{figure5}(a), the transparency of the four-level model (red line) is enhanced with respect to the one for the two three-level model for low control field Rabi frequencies. As explained in the Sect.~\ref{TheorM:A}, see Fig.~\ref{Figure2}, for the low Rabi frequency limit  Raman absorption due to the response of $\chi_{42}$  occurs on two-photon resonance for the two three-level model while it is red shifted for the four-level model. This shift, in combination with the gain from the response of $\chi_{42}$ for the four-level model, leads to the enhanced transparency for the four-level model.  As expected, the transparency for both models increases as the Rabi frequency increases. However, the transparency for the four-level model saturates below one, while the one for the two three-level model tends to a value of one. This results from the fact that for the four-level model the red shifted EIT is always limited by the off-resonance Raman absorption. On the other hand, for the two three-level model the EIT transmission does not shift while the Raman absorption shifts as the control field Rabi frequency is increased.

We next consider the effect of the ground state decoherence rate on the transparency, see Fig.~\ref{figure5}(c). While  perfect transparency can be achieved for the three-level model, the transparency for the four-level model never reaches a value of one. This is the case even in the ideal limit of no decoherence. This result implies that it is not possible to obtain complete transparency in the D1 line of alkali atoms which have unequal dipole moments for the involved transitions in hot vapor cells where Doppler broadening needs to be taken into account. This result is consistent with the one in~\cite{EITGiacobino1}, where they showed that the presence of multiple excited levels significantly reduces the EIT transmission when Doppler broadening is present.

Given that the width of the EIT transmission peak affects the dispersion in the medium, it has an impact on the group velocity of the light~\cite{EITREVTittel,StorageIrina,EITRonald}.  It is thus important to develop a more accurate description of the response of the atomic system to understand the effect of the two closely-spaced excited states.  To do this, we study the effect of control field power and decoherence on the EIT width for the four-level and the two three-level models, as shown in Figs.~\ref{figure5}(b) and~\ref{figure5}(d).   For both models, the FWHM increases linearly as the control field Rabi frequency and decoherence rate increase, following the behavior that has been previously predicted using a three-level calculation~\cite{EITBretenaker,EITLinearLvovsky,EITlinewidthScully}. As can be seen from Fig.~\ref{figure5}(b), the EIT transmission peak for the four-level model is broader than the one for the three-level model.  This is a result of the gain (in the response due to $\chi_{42}$ for the four-level model in Fig.~\ref{Figure2}) to the blue of the EIT peak, which broadens with control power and contributes to the EIT width. For high decoherence rates, the gain and the Raman absorption in the response of the four-level model slowly vanish. As a result, the four-level and two three-level models converge to the same FWHM.

\subsection{\label{TheorM:C}Role of separation between excited states}

\begin{figure}[htb]
  \centering
  \includegraphics{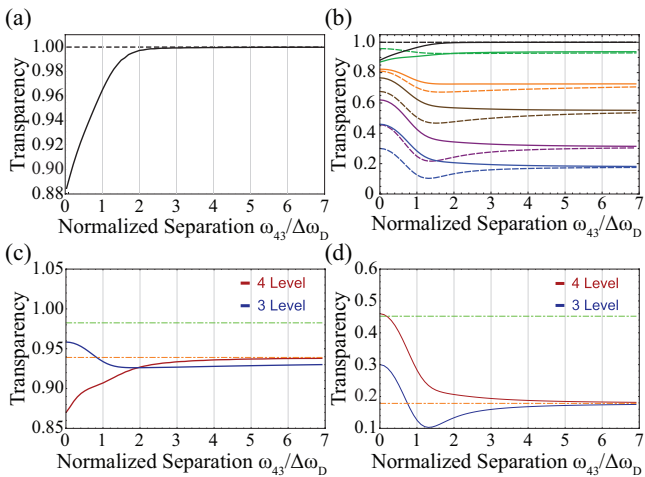}
  \caption{EIT transparency for the Doppler broadened atomic response as a function of the normalized frequency separation between the excited states. (a) EIT transparency for the ideal case of no decoherence, $\gamma_{21}$=0, for different temperatures ($T=0^{\circ}$C, 100$^{\circ}$C, 200$^{\circ}$C, 300$^{\circ}$C, and 400$^{\circ}$C). (b) EIT transparency at $T=50^{\circ}$C for different decoherence rates [$\gamma_{21}=0$ (black), 0.01$\gamma$ (green), 0.05$\gamma$ (orange), 0.1$\gamma$ (brown), 0.25$\gamma$ (purple), and 0.5$\gamma$ (blue)]. The response for the four-level and the three-level models are shown with solid and dashed lines, respectively. Figures (c) and (d) show a comparison of the EIT transmission at $T=50^{\circ}$C for the four-level and two-three level models with the ones of simple three level systems for decoherence rates of $\gamma_{21}=0.01\gamma$ and $\gamma_{21}=0.5\gamma$, respectively. The dashed-dotted lines represent the EIT transmission for a single-lambda system at one-photon resonance for an effective dipole moment for the control field transition of $d_{31}$ (orange) and $d_{41}$ (green). The calculations are done for $\Delta=0$, $\Omega_{31}=5\gamma$, and a cell length of 2.54~cm.}
  \label{figure6}
\end{figure}

To understand the role of the two closely-spaced excited levels on EIT, we consider the effect of the frequency separation between them on the transparency. As was shown in Sect.~\ref{TheorM:B}, it is not possible to obtain  perfect transparency when both excited levels are taken into account in the presence of Doppler broadening even for the ideal case of no decoherence. We revisit this point and consider the effect of the spacing between the excited levels. To do so, we calculate the Doppler broadened EIT transparency for zero decoherence as a function of the normalized separation, which we define as the ratio of the frequency separation between the excited states and the Doppler width, as shown in Fig.~\ref{figure6}(a).

For the four-level model, which captures the response from an effective coupling between the two excited states, we find that the transparency depends only on the normalized separation and not the absolute frequency separation between the excited levels.  This can be seen from Fig.~\ref{figure6}(a), which shows that the behavior of the transparency is independent of the temperature ($T=20^{\circ}$C, 50$^{\circ}$C, 100$^{\circ}$C, 200$^{\circ}$C, and 400$^{\circ}$C), which effectively changes the Doppler broadened width. For reference, the figure also shows the result from the two three-level model (black dashed line), which shows complete transparency independent of the separation between the excited states. As can be seen, when it is not possible to clearly resolve the two excited levels ($\omega_{43}/\Delta\omega_{D}<2$) complete transparency is not possible. It is however possible to recover full transparency when the frequency separation between the excited levels is more than twice the Doppler width.

In Fig.~\ref{figure6}(b) we explore the behavior of the transparency with respect to the normalized separation for multiple decoherence rates [$\gamma_{21}=0$ (black), 0.01$\gamma$ (green), 0.05$\gamma$ (orange), 0.1$\gamma$ (brown), 0.25$\gamma$ (purple), and 0.5$\gamma$ (blue)] at a temperature of $T=50^{\circ}$C. The results from the two three-level model are shown with dashed lines, while the ones for the four-level model are shown with solid lines. We find that as the decoherence rate increases, the transparency of the four-level model is enhanced relative to the one of the two three-level model, in particular in the regime where it is not possible to clearly resolve the two excited levels ($\omega_{43}/\Delta\omega_{D}<2$). This implies that EIT in a vapor cell is enhanced for a small normalized separation between the two excited states. For a large normalized separation, both models converge to the same transmission level, with the transparency for the four-level model always higher than the one for the three-level model. This can be understood by referring to Fig.~\ref{Figure2}. As the normalized separation increases, level $\ket{4}$ moves further and further away from resonance, which for the four-level model leads to the reduction of the gain to the blue of the EIT peak (which enhanced the transparency), while for the two-three level model this leads to the reduction of the off-resonance Raman absorption (which suppressed the transparency).

To obtain a better understanding of the effect of the two excited states on EIT, we consider the suppression and enhancement of the transparency with respect to a simple three-level system for decoherence rates of 0.01$\gamma$ and 0.5$\gamma$ in Figs.~\ref{figure6}(c) and~\ref{figure6}(d), respectively. In particular, we compare the transparency of the four-level and two-three level models with the ones of the two independent three-level systems that make up the two three-level model. The dashed-dotted lines represent the transparency for a simple three-level system with a dipole moment for the control field transition of $d_{41}$ (green) and $d_{31}$ (orange) for both control and probe fields on resonance with their corresponding transition (effectively for a normalized separation of zero). Interestingly, the transparency of the four-level model for zero normalized separation is lower than that of both simple three-level systems for the case of low decoherence rate, see Fig.~\ref{figure6}(c), and larger than that of both simple three-level systems for a larger decoherence rate, see Fig.~\ref{figure6}(d). This suggests that the EIT transmission enhancement and suppression cannot be explained with a single three-level lambda system and that they are a result of the coherent coupling of the control and probe fields with the two excited levels. From Figs.~\ref{figure6}(c) and~\ref{figure6}(d) we can also see that in the limit of a large normalized separation, the transparency tends to the one of a simple three-level system with dipole moment $d_{31}$ (orange dashed-dotted line). As the normalized separation increases energy level $\ket{4}$ moves further away from resonance, which reduces the effect of the lambda system with $d_{41}$ on the EIT response of the atom.

\subsection{\label{TheorM:D}Off-resonance Raman}

\begin{figure}[htb]
  \centering
  \includegraphics{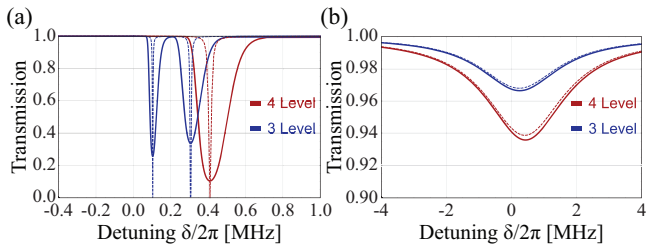}
  \caption{Transmission for the off-resonance Raman process as a function of the two-photon detuning $\delta$/2$\pi$ with (solid) and without (dashed) Doppler broadening  for (a) $\gamma_{21}$=0 and (b) $\gamma_{21}=0.25\gamma$ (see Sect.~\ref{ExperiR:A}). The calculations are done for $T=50^{\circ}$C, $\Omega_{31}=5\gamma$, $\Delta=2$~GHz, and a cell length of 2.54~cm.}
  \label{Figure10}
\end{figure}

The two closely-spaced excited states also play a role in the off-resonance Raman process where the fields are tuned far away from resonance (large one-photon detuning). For the three-level model, each of the three level systems will exhibit its own off-resonance Raman absorption. This behavior can be seen in Fig.~\ref{Figure10}(a) (blue trace) where two absorption dips, corresponding to  Raman transitions through levels $\ket{3}$ and $\ket{4}$, are clearly visible. Since the detuning of the control field is different for each of the two excited levels due to the frequency separation between them, each of the Raman processes experiences a different light shift.  This leads to the off-resonance Raman absorption dips associated with levels $\ket{3}$ and $\ket{4}$ to appear at different frequencies, around $\delta/2\pi=0.1$~MHz and $\delta/2\pi=0.3$~MHz, respectively. These locations are consistent with the expected light shifts, $\delta_{3}^{\ket{3}}=\Omega_{31}^2/4\Delta$ for the one associated with level $\ket{3}$ and $\delta_{3}^{\ket{4}}=\Omega_{41}^2/4(\Delta+\omega_{43})$ for the one associated with level $\ket{4}$. On the other hand, the four-level model, see red line in Fig.~\ref{Figure10}(a), shows only one Raman absorption dip near $\delta/2\pi=0.4$~MHz, which means that the two excited levels need to be treated as an effective single excited level. In addition, the interplay between the two excited states leads to an enhanced off-resonance Raman absorption and a more significant light shift, which is now approximately equal to the sum of the light shifts for each of the two individual lambda systems,  $\delta_{4}\approx\delta_{3}^{\ket{3}}+\delta_{3}^{\ket{4}}$.

Similar to the results we found for EIT, we find that for off-resonance Raman absorption the  four-level model exhibits enhanced absorption with a broader lineshape for larger decoherence rates, as shown in Fig.~\ref{Figure10}(b). Since off-resonance Raman transitions can be used in an atomic vapor to implement a quantum memory~\cite{EIToffresBuchler} and can impact atomic interferometers though phase shifts~\cite{EIToffresLandragin}, the slight frequency shift of the resonance and its increased absorption level and width will lead to a slightly different phase shift and storage time from what would be expected from a simple three-level system. Thus, these effects need to be properly taken into account when working with real atomic systems.

\subsection{\label{TheorM:E}Dressed state picture}

In order to obtain additional insight into the physics of the results presented above, we now consider a dressed state model. Detailed calculations of the eigenvalues of the dressed atomic system are given in Appendix~\ref{app:C}. Figure~\ref{figure7} shows a contour plot of the transmission spectrum obtained from the response of the calculated susceptibilities as a function of the two-photon detuning $\delta$ and the one-photon detuning $\Delta$. The eigenvalues of the dressed state model, shown as magenta dashed line, show the effective location of the resonances for the probe beam~\cite{DressReynaud}. As can be seen, the eigenvalues coincide with the locations of maximum absorption obtained from a full calculation of the atomic response. As a result, this approach provides a simple way to calculate the location of the EIT resonance and obtain a better understanding of how different parameters impact the transparency.
\begin{figure}[htb]
  \centering
  \includegraphics{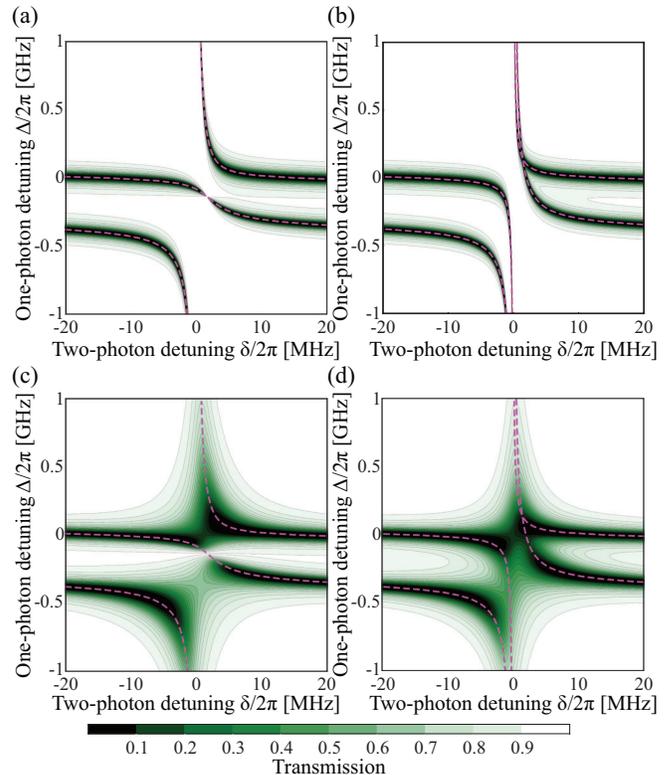}
  \caption{Contour plots of the transmission spectra as a function of the two-photon detuning $\delta/2\pi$ and the one-photon detuning $\Delta/2\pi$. Transmission spectrum for (a) the four-level model with $\gamma_{21}=0$, (b) the two three-level model with $\gamma_{21}=0$, (c) the four-level model with $\gamma_{21}=0.25\gamma$ (see Sect.~\ref{ExperiR:A}), and (d)~the two three-level model with $\gamma_{21}=0.25\gamma$ (see Sect.~\ref{ExperiR:A}). The eigenvalues obtained from the dressed state model are represented with the magenta dashed lines. The calculations are done for $T=50^{\circ}$C, $\Omega_{31}=5\gamma$, and a cell length of 2.54~cm.}
  \label{figure7}
\end{figure}

As can be seen in Fig.~\ref{figure7}(b), for the two three-level model the two eigenvalues diverge near the two-photon resonance.  As a result, at exact two-photon resonance there is complete transparency independent of the one-photon detuning~\cite{EITRevMarangos}. The two diverging eigenvalues correspond to the two off-resonance Raman process shown in Sect.~\ref{TheorM:D} for large one-photon detuning $\Delta$.
On the other hand, for the four-level model, shown in Fig.~\ref{figure7}(a), one of the eigenvalues does not exhibit such a divergence.  As a result, perfect transparency is in general not present on two-photon resonance. A small transparent region appears to the red of the two-photon resonance, which has been previously observed~\cite{QunmIntZheng,flevelGao}. The location of this transparency region in terms of the one-photon and two-photon detuning can be calculated using the dressed state solution and is given by the location where the transition probability amplitudes of the transitions $\ket{3}$ to $\ket{1}$ and $\ket{4}$ to $\ket{1}$ have equal and opposite magnitudes. The absorption probability vanishes when
\begin{align}
\Delta_\textrm{vanish} =
   \frac{4\,\omega_{43}^2 \Omega_{31}+(\Omega_{31}-\Omega_{41}) (\Omega_{31}+\Omega_{41})^2}{4\,\omega_{43}(\Omega_{31}+\Omega_{41})},\label{eq1}
\end{align}

\begin{align}
\delta_\textrm{vanish} =
   \frac{\Omega_{31}^2-\Omega_{41}^2}{4\,\omega_{43}},\label{eq2}
\end{align}

\noindent which leads to complete transparency.

As can be seen from these results, for transitions coupled by the control field with equal dipole moments ($d_{41}=d_{31}$), and thus equal Rabi frequencies ($\Omega_{41}=\Omega_{31}$), $\Delta_\textrm{vanish}$ becomes half the separation between the two excited states and $\delta_\textrm{vanish}$ becomes zero. As a result, complete transparency is obtained on two-photon resonance when the control field is tuned half way between the two excited levels. In this case complete EIT transparency is also present even when Doppler broadening is taken into account, as Doppler broadening results in a broadening mainly along the direction of the one-photon detuning $\Delta$ (vertical axis in Fig.~\ref{figure7}) for co-propagating control and probe fields. For the case of different dipole strengths, the location of maximum transparency is shifted off the two-photon resonance.  Thus, when Doppler broadening is taken into account, the upper branch due to the other excited state in Fig.~\ref{figure7}(a) starts to merge with the transparency region if the separation between the two excited levels is smaller than the Doppler width.  As a result, it is not possible to obtain perfect transparency even in the ideal case of no decoherence unless the separation between the excited levels is larger than the Doppler broadening.

The shift of the two-photon resonance, shown in Fig.~\ref{figure7}(a), will also create an asymmetric behavior for the two possible EIT conditions at $\Delta=0$ and $\Delta=-\omega_{43}$. Since the central continuous branch from the dressed state picture moves towards the corresponding branch associate with level $\ket{3}$, there will be a larger reduction of the EIT transparency when the resonant excited state for the EIT is level $\ket{3}$ than when it is level $\ket{4}$. This will cause the EIT associated with level $\ket{4}$ to have a relatively larger transparency level than the EIT associated with level $\ket{3}$. As a result of such an asymmetry, an increase in separation between the two excited states or a large imbalance in dipole moments of the transitions will lead to larger differences between EIT associated with the upper and lower excited level.

The behavior changes in the limit of large decoherence rates, see Figs.~\ref{figure7}(c) and (d). While for this limit the location of the EIT resonances stays the same, the additional decoherence leads to absorption in the region around the two-photon resonance where there was previously transparency due to EIT. It is interesting to note that in this limit there is still a region of perfect transparency for the four-level model while this is not the case for the two-independent three-level model. Once Doppler broadening is taken into account this is no longer the case; however, it leads to a larger transparency for the four-level system than for the two-three level systems, as previously shown in Fig.~\ref{figure5}(c).

\section{\label{ExperiR}Experimental Results}
\begin{figure}[t!]
  \centering
  \includegraphics{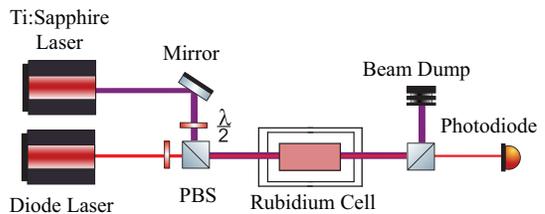}
  \caption{Simplified schematic of the experimental setup. The pump and probe beams co-propagate inside an isotropically pure $^{85}$Rb or $^{87}$Rb vapor cell. After the cell, the pump beam is filtered with a PBS and the intensity of the probe is measured with a photodiode. PBS: polarizing beam splitter.}
  \label{figure8}
\end{figure}

In this section we present experimental measurements of EIT in the D1 lines of $^{85}$Rb and $^{87}$Rb to compare with the theoretical calculations from the four-level model. We use a CW Ti:Sapphire laser tuned to the corresponding D1 line at around 795 nm to generate the control beam and a home-built external cavity diode laser system for the probe beam,  as shown in Fig.~\ref{figure8}. The frequency difference between the lasers is set to the corresponding frequency separation of the two ground state hyperfine levels of $^{85}$Rb or $^{87}$Rb. The control and probe fields are both sent through optical fibers to clean up their spatial profiles. After the fibers, optical systems are used to obtain $1/e^2$ diameters for the control and probe fields of 0.9~mm and 0.75~mm, respectively. Perpendicularly polarized control and probe beams are combined with a polarizing beam splitter and sent in a co-propagating configuration through a 3 inch isotropically pure $^{85}$Rb or $^{87}$Rb vapor cell with two layers of magnetic shielding. After the cell, the control beam is blocked with a polarization filter and the probe is measured with a photodiode to obtain the transmission spectrum. Both cells are slightly heated to obtain similar absorption spectra for both Rb isotopes.  We then extract the temperatures of the cells by fitting each absorption spectrum to the four-level calculation in the absence of the control field. This leads to estimated temperatures for the $^{85}$Rb and $^{87}$Rb cells of 24.5$^{\circ}$C and 28$^{\circ}$C, respectively.

\subsection{\label{ExperiR:A}EIT transmission spectra}

\begin{figure}[t!]
  \centering
  \includegraphics{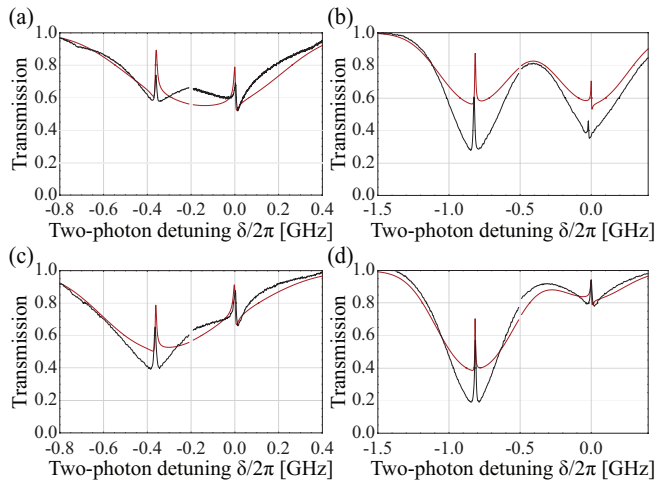}
  \caption{Experimental EIT transmission spectra in the D1 line of $^{85}$Rb and $^{87}$Rb as a function of the two-photon detuning $\delta/2\pi$. The combined measured EIT spectra (black traces) and the theoretical predictions from the four-level model (red traces) are shown for cases in which the control field is on resonance with the (a) $F=2$ to $F'$ transition of $^{85}$Rb, (b) $F=1$ to $F'$ transition of $^{87}$Rb, (c) $F=3$ to $F'$ transition of $^{85}$Rb, and (d) $F=2$ to $F'$ transition of $^{87}$Rb. The length of both the $^{85}$Rb and $^{87}$Rb vapor cells is $\sim7.6$~cm, while their temperatures are set to $T_{85}=24.5^{\circ}$C and $T_{87}=28^{\circ}$C, respectively. For all cases, the power of control field is set to 15~mW with a $1/e^{2}$ diameter of 0.9~mm. The decoherence rate between the two ground states is estimated to be $\gamma_{21}=0.25\gamma$ by simultaneously optimizing the fit to all possible EIT transitions.}
  \label{figure9}
\end{figure}

We first present experimental results for EIT for all possible transitions in the D1 lines of both Rb isotopes, as shown in Fig.~\ref{figure9}. Each plot contains two separate EIT spectra with the control field set to resonance with one or the other excited state. The transmission spectra for the theoretical predictions from the four-level model and the experimental results are shown with red and black traces, respectively, as a function of the two-photon detuning.
Figure~\ref{figure9}(a) shows, for example, the EIT spectra when the control field is set to resonance with the D1 $F=2$ to $F'=2$ transition (right) and the D1 $F=2$ to $F'=3$ transition (left) of $^{85}$Rb.  These two spectra were measured independently and then combined in a single figure, with a discontinuity at $\delta/2\pi=0.2$. As can be seen, in general, there is good agreement between the theoretical predictions and the experimental results.

It is worth noting, however, that there are limitations in the theoretical model that lead to a mismatching between the experimental and theoretical results. First, our analytical model assumes that the control field is in the strong limit, which would lead to perfect optical pumping (all the population in one of the ground states). Therefore, the theoretical predictions shown in Fig.~\ref{figure9} exhibit higher transmission and broader absorption spectra than the measured spectra. Second, our model does not allow the coupling of the control field to the $\ket{2}\rightarrow\ket{3},\ket{4}$ transitions. Such a coupling would lead to a light shift of the transition and the possibility four-wave mixing process in a double-$\Lambda$ configuration~\cite{Wong03}. Third, we are not considering the contributions from the magnetic sublevels, which have been shown to modify the behavior of EIT and lead to electromagnetically induced absorption (EIA)~\cite{Lezama99a,Lezama99b}. The mismatch in absorption near the conditions for EIT is due to the contribution from the EIA. In addition to the limitations of the model, the comparison between the theory and experiment requires the estimation of the number density, effective Rabi frequencies, and decoherence rates. In particular, the decoherence rate is estimated by simultaneously optimizing the fit to the EIT transmission spectra for all possible transitions shown in Fig.~\ref{figure9}, which gives an estimated value of $\gamma_{21}=0.25\gamma$. The decoherence rate is most likely dominated by the use of two independent lasers for the probe and conjugate fields.

\subsection{\label{ExperiR:B}Dependence of EIT on Rabi frequency}

\begin{figure}[t!]
  \centering
  \includegraphics{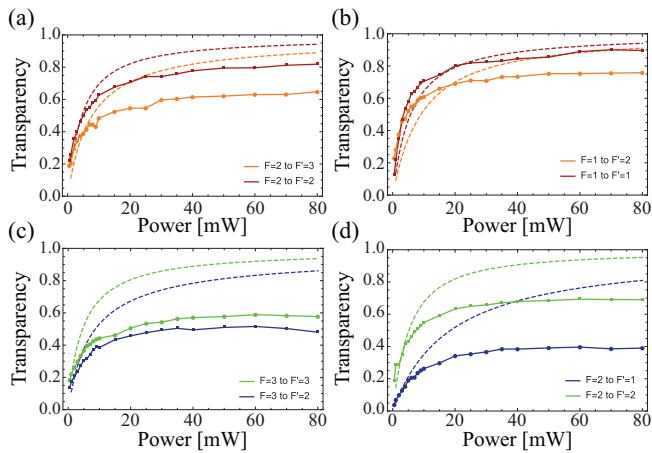}
  \caption{Experimental results for the EIT transparency in the D1 transition of $^{85}$Rb and $^{87}$Rb as a function of the Rabi frequency of the control field. The measured EIT transparencies (solid line) and the theoretical predictions from the four-level model (dashed lines) are shown for the control field on resonance with the (a) $F=2$ to $F'$ transition of $^{85}$Rb, (b) $F=1$ to $F'$ transition of $^{87}$Rb, (c) $F=3$ to $F'$ transition of $^{85}$Rb, and (d) $F=2$ to $F'$ transition of $^{87}$Rb. The length of both the $^{85}$Rb and $^{87}$Rb vapor cells is $\sim7.6$~cm, while their temperatures are set to $T_{85}=24.5^{\circ}$C and $T_{87}=28^{\circ}$C, respectively. The $1/e^{2}$ diameter of the control field is 0.9~mm. The estimated decoherence rate is $\gamma_{21}=0.25\gamma$.}
  \label{figure10}
\end{figure}

We next compare the experimental results and the theoretical predictions from the four-level model for the dependence of the EIT transparency on the Rabi frequency of the control field, as shown in Fig.~\ref{figure10}. To calculate the transparency, as defined in Sect.~\ref{TheorM:B}, from the experiment experimental data, we extract $\chi_\textrm{EIT}$ from the peak value of the EIT and estimate $\chi_\textrm{abs}$ from the minimum transmission value near the EIT peak. Note that this value will deviate from the one given by the theory model due to the presence of EIA, as explained in the previous section.

In the low control field regime, the experimental data and theoretical predictions show good agreement. Furthermore, the relative behavior of the transparencies for all the possible transitions over the whole power range is consistent between theory and experiment. In general we observe a common behavior from the experimental results: EIT when the control field couples to the lower ground state, shown in Figs.~\ref{figure10}(a) and~\ref{figure10}(b), shows higher transparency when coupling to the lower excited state. On the other hand, EIT when the control field couples to the upper ground state, shown in Figs.~\ref{figure10}(c) and~\ref{figure10}(d), shows lower transparency when coupling to the lower excited state.

There are two main reasons for this behavior. First, the transitions have different dipole moments, which leads to different Rabi frequencies. For example, in $^{85}$Rb the Rabi frequency for the D1 $F=2$ to $F'=3$ transition is larger than the one for the D1 $F=2$ to $F'=2$ transition for the same optical power given that $d_{F=2 \rightarrow F'=3}>d_{F=2 \rightarrow F'=2}$, but the Rabi frequency of the D1 $F=3$ to $F'=3$ transition is lower than the one for the D1 $F=3$ to $F'=2$ transition for the same optical power given that $d_{F=3 \rightarrow F'=3}<d_{F=3 \rightarrow F'=2}$. Second, as explained in Sect.~\ref{TheorM:E}, transitions with unequal dipole moments lead to a shift of the two-photon detuning at which maximum EIT transparency happens in a direction towards the transition with the weaker dipole moment strength. This shift makes the transparency regions around each transition asymmetric, which enhances the EIT for the transition with the higher dipole strength and decreases the EIT for the transition with the lower dipole strength.

These results also demonstrate the effect of the separation between the two excited states by comparing the data for $^{85}$Rb and $^{87}$Rb (with excited state frequency separation $\omega_{43}^{(85)}=2\pi\times361$~MHz and $\omega_{43}^{(87)}=2\pi\times817$~MHz, respectively) given that the dipole strengths ($d_{FF'}$) for the different transitions in the two isotopes are relatively close to each other  ($d_{23}^{(85)}$$\approx$$d_{12}^{(87)}$, $d_{22}^{(85)}$$\approx$$d_{11}^{(87)}$, $d_{32}^{(85)}$$\approx$$d_{21}^{(87)}$, and $d_{33}^{(85)}$$\approx$$d_{22}^{(87)}$).  Since the dipole strengths are similar to each other but the separation between the two excited states is substantially larger in $^{87}$Rb, the EIT transparency region will be in general shifted from $\delta=0$ and $\Delta=0$. From Eqs.~(\ref{eq1}) and~(\ref{eq2}) we can see that the shifts will be larger for $^{87}$Rb than for $^{85}$Rb. This implies that we should observe a larger difference between the transparencies in $^{87}$Rb than for the corresponding transitions in $^{85}$Rb. This behavior can be seen in Figs.~\ref{figure10}(c) and~\ref{figure10}(d), in which the difference between transparencies in the two possible transitions in $^{87}$Rb is larger than the difference in $^{85}$Rb.

Interestingly, we observe the opposite behavior in Figs.~\ref{figure10}(a) and~\ref{figure10}(b). This is because the two dipole moments for the transitions in $^{87}$Rb ($d_{21}^{(87)}$ and $d_{22}^{(87)}$) are the exactly same, while the ones for $^{85}$Rb ($d_{22}^{(85)}$ and $d_{32}^{(85)}$) are still slightly different. This leads to smaller shifts of the transparent region for $^{87}$Rb and thus a more symmetric situation. Therefore, EIT in $^{85}$Rb, shown in Fig.~\ref{figure10}(a), exhibits a larger difference in transparency between the two transitions than between the corresponding transitions in $^{87}$Rb, shown in Fig.~\ref{figure10}(b). Furthermore, given that the dipole strengths in $^{87}$Rb are exactly the same, higher transparency can in theory be achieved, as experimentally verified by comparing Figs.~\ref{figure10}(a) and~\ref{figure10}(b).

\subsection{\label{ExperiR:C}Dependence of EIT on one- and two-photon detunings}

\begin{figure}[t!]
  \centering
  \includegraphics[width=\columnwidth]{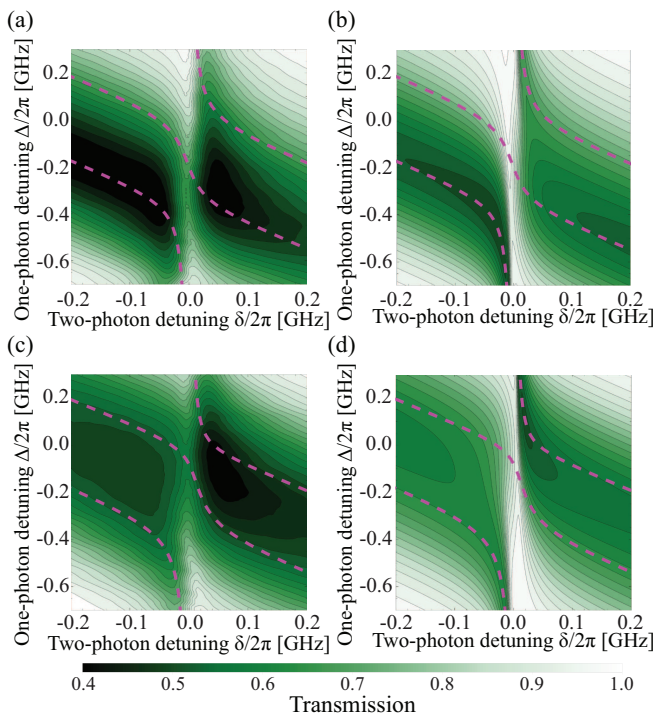}
  \caption{Contour plots of the normalized transmission spectra as a function of the two-photon detuning $\delta/2\pi$ and the one-photon detuning $\Delta/2\pi$ for the experimental measurements of EIT with the control field on resonance with (a) the D1 $F=2$ to $F'$ transition and (c) D1 $F=3$ to $F'$ transition of $^{85}$Rb, and the theoretical calculations of the Doppler broadened transmission for the four-level model with the control field on resonance with the (b) the D1 $F=2$ to $F'$ transition and (d) D1 $F=3$ to $F'$ transition of $^{85}$Rb. The eigenvalues obtained from the dressed state model are represented by the magenta dashed lines. For both the experimental and theoretical figures, the parameters are given by $T=24.5^{\circ}$C, control field power and $1/e^{2}$ diameter of 80~mW and 0.9~mm, respectively, and a cell length of 7.6~cm. The estimated decoherence rate is $\gamma_{21}=0.25\gamma$.}
  \label{figure11}
\end{figure}

\begin{figure}[t!]
  \centering
  \includegraphics[width=\columnwidth]{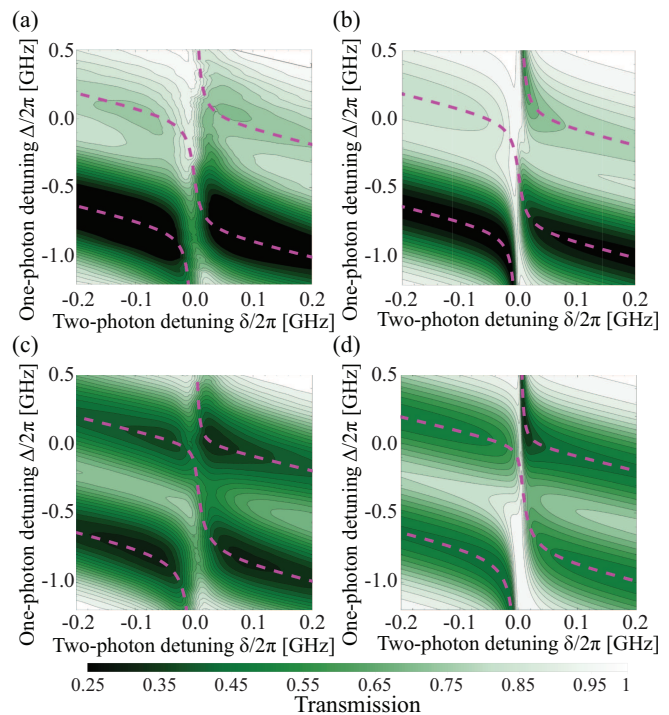}
  \caption{Contour plots of the normalized transmission spectra as a function of the two-photon detuning $\delta/2\pi$ and the one-photon detuning $\Delta/2\pi$ for the experimental measurements of EIT with the control field on resonance with (a) the D1 $F=1$ to $F'$ transition and (c) D1 $F=2$ to $F'$ transition of $^{87}$Rb, and the theoretical calculations of the Doppler broadened transmission for the four-level model with the control field on resonance with (b) the D1 $F=1$ to $F'$ transition and (d) D1 $F=2$ to $F'$ transition of $^{87}$Rb. The eigenvalues obtained from the dressed state model are represented by the magenta dashed lines. For both the experimental and theoretical figures, the parameters are given by $T=28^{\circ}$C, control field power and $1/e^{2}$ diameter of 80~mW and 0.9~mm, respectively, and a cell length of 7.6~cm. The estimated decoherence rate is $\gamma_{21}=0.25\gamma$.}
  \label{figure12}
\end{figure}

To better compare with the behavior seen in the dressed state model, we measure the EIT transmission spectra for the probe beam as we scan both the one- and two-photon detunings around the D1 lines of $^{85}$Rb and $^{87}$Rb. Figures~\ref{figure11} and~\ref{figure12} show the contour plots of the normalized transmission spectra for both the experimental results and the theoretical calculations obtained with the four-level model for $^{85}$Rb and $^{87}$Rb, respectively. In both figures, the experimental results are shown on the left column and the theoretical predictions are shown on the right column. The magenta dashed lines correspond to the eigenvalues obtained from the dressed state model. As can be seen, the experimental data shows the same behavior as the theoretical model and agrees well with the location of the resonances from the dressed state model. In particular, both the theory and the data show a similar diverging behavior as the one-photon detuning increases (off-resonant Raman transition) near the two-photon resonance.

In Figs.~\ref{figure11}(a) and ~\ref{figure11}(b), the one-photon resonance $\Delta=0$ and two-photon resonance $\delta=0$ correspond to a control field resonant with the D1 $F=3$ to $F'=2$ transition and a probe field resonant with the D1 $F=2$ to $F'=2$ of $^{85}$Rb. Given that the dipole moments for the D1 $F=3$ to $F'=2$ and D1 $F=3$ to $F'=3$ transitions the control field can couple to are almost the same, the eigenvalues from the dressed state model are equally separated from the two-photon detuning at which transparency is achieved.  This behavior is consistent with both the experimental results, see Fig.~\ref{figure11}(a), and the theoretical model, see Fig.~\ref{figure11}(b). One the other hand, for Figures~\ref{figure11}(c) and ~\ref{figure11}(d), for which $\Delta=0$ and $\delta=0$ correspond to a control field resonant with the D1 $F=2$ to $F'=2$ transition and a probe field resonant with the D1 $F=3$ to $F'=2$ of $^{85}$Rb, the dipole moments for the transitions $F=2$ to $F'=2$ and $F=2$ to $F'=3$ are relatively large.  As a result, two of the eigenvalues from the dressed state model tend to get closer at $\Delta=0$ and $\delta=0$, which limits the EIT transmission at $\Delta=0$ but increases the transmission at $\Delta=-351$~MHz.

Similar behavior can be seen for the different transitions in $^{87}$Rb, as shown in Fig.~\ref{figure12}.  It is worth noting, however, that Figs.~\ref{figure12}(a) and ~\ref{figure12}(b) show higher EIT transmission than the corresponding configuration in $^{85}$Rb, Figs.~\ref{figure11}(a) and ~\ref{figure11}(b). As discussed in Sects.~\ref{TheorM:E} and~\ref{ExperiR:B}, the combination of having the dipole moments for the D1 $F=2$ to $F'=1$ and D1 $F=2$ to $F'=2$ transitions in $^{87}$Rb be the same and a larger separation between the excited state leads to such an increase in EIT transparency.

\section{\label{Conl}Conclusion}
We present a study of the effects of two closely-spaced excited states on EIT and off-resonance Raman transitions. To identify the effects due to the coherent interaction of the fields with the two excited states we compare two models, one which includes the four levels and one that includes two independent three-level systems. We use these two models to study the behavior of EIT and off-resonance Raman as a function of the decoherence rate, Rabi frequency, and the separation between the two excited levels.

We find that for high decoherence rates the presence of two closely-spaced excited states whose frequency separation is less than their Doppler broadened linewidths enhances the EIT transmission and broadens the width of the EIT resonance. However, even in the ideal case of no decoherence perfect transparency is not possible and can only be recovered when the separation between the excited states is larger than the Doppler broadening or when the dipole moments are the same for the two possible transitions for the control field. These effects can play a role on quantum devices based on EIT in a room temperature vapor cell as a broader transparency region with reduced transmission introduces losses and limits the amount of group delay that can be achieved~\cite{EITREVTittel}. We also show that the presence of two closely-spaced excited states enhances off-resonance Raman transitions and introduces a larger light shift.

Finally, we present a dressed state model to explain the observed behaviors and perform measurements of EIT in the D1 lines of $^{85}$Rb and $^{87}$Rb to compare with the theoretical predictions. Our measured transmission spectra agrees well with the theoretical calculations based on the four-level model and the eigenvalues from the dressed state picture. This analysis provides a physical picture that successfully explains the role of two closely-spaced excited states on the behavior of EIT in all the possible transitions in the D1 line of Rb.

\begin{acknowledgments}
We would like to thank Kit A. Leonard for providing support with the external cavity diode laser needed for the experiment. This work was supported by a grant from AFOSR (FA9550-15-1-0402).
\end{acknowledgments}

\appendix

\section{\label{app:A}Derivation of susceptibility}

We use the density matrix formalism to derive the atomic response of the four-level system. In order to derive the susceptibilities for the two transitions that the probe beam can couple to, we find the equations of motion for the density matrix elements and solve them in steady state.  After applying the rotating wave approximation, the equations of motion for the density matrix elements take the form
\begin{align}
\dot{\sigma_{43}}&=-(i \omega_{43}+\gamma_{43})\sigma_{43}\\
      &\qquad+\frac{i}{2}(\Omega_{41}\sigma_{13}+\Omega_{42}\sigma_{23}-\Omega_{13}\sigma_{41}-\Omega_{23}\sigma_{42})\nonumber,
\end{align}
\begin{align}
\dot{\sigma_{42}}&=-(i \omega_{43}+\gamma_{42}+\delta+\Delta)\sigma_{42}\\
      &\qquad+\frac{i}{2} (\Omega_{41}\sigma_{12}+\Omega_{42}\sigma_{22}-\Omega_{32}\sigma_{43}-\Omega_{42}\sigma_{44})\nonumber,
\end{align}
\begin{align}
\dot{\sigma_{41}}&=-(i \omega_{43}+\gamma_{41}+i\Delta)\sigma_{41}\\
      &\qquad+\frac{i}{2} (\Omega_{41}\sigma_{11}+\Omega_{42}\sigma_{21}-\Omega_{31}\sigma_{43}-\Omega_{41}\sigma_{44})\nonumber,
\end{align}
\begin{align}
\dot{\sigma_{32}}&=-(\gamma_{32}+i\Delta+i\delta)\sigma_{32}\\
      &\qquad+\frac{i}{2} (\Omega_{31}\sigma_{12}+\Omega_{32}\sigma_{22}-\Omega_{32}\sigma_{33}-\Omega_{42}\sigma_{34})\nonumber,
\end{align}
\begin{align}
\dot{\sigma_{31}}&=-(\gamma_{31}+i\Delta)\sigma_{31}\\
      &\qquad+\frac{i}{2} (\Omega_{31}\sigma_{11}+\Omega_{32}\sigma_{21}-\Omega_{31}\sigma_{33}-\Omega_{41}\sigma_{34})\nonumber,
\end{align}
\begin{align}
\dot{\sigma_{21}}&=-(\gamma_{21}-i \delta)\sigma_{21}\\
      &\qquad+\frac{i}{2} (\Omega_{32}\sigma_{31}+\Omega_{42}\sigma_{41}-\Omega_{31}\sigma_{23}-\Omega_{41}\sigma_{24})\nonumber,
\end{align}
\noindent where $\sigma_{ij}$ are the density matrix elements after the rotating wave approximation, $\gamma_{ij}$ are the dipole dephasing rates, $\Omega_{ij}$ are Rabi frequencies for the transition between levels $\ket{i}$ and $\ket{j}$, and $\omega_{43}$ is the frequency difference between the two excited states.

We analytically solve this system of equations in steady state for the off-diagonal elements of the density matrix. For simplicity we rewrite the Rabi frequencies as $\Omega_{41}=\alpha~\Omega_{31}$ and $\Omega_{42}=\beta~\Omega_{32}$, where $\alpha=d_{41}/d_{31}$ and $\beta=d_{42}/d_{32}$ with $d_{ij}$ representing the dipole moment of the corresponding  transitions. For the case of $^{85}$Rb $\alpha=\sqrt{7/2}$ and $\beta=\sqrt{4/5}$. For a weak probe beam, the strong pump optically pumps the atomic system to level $\ket{2}$. After taking into account that all the population is in $\ket{2}$, we find that in the weak probe approximation the steady state solution for the density matrix elements for the transitions that the probe can couple to are given by
\begin{align}
\sigma_{42}&=-\frac{(\alpha-\beta)|\Omega|^2+4\beta(\delta -i \gamma_{21})(-i \gamma_{32}+\delta +\Delta)}{Z}\Omega_{32},\label{sigma42}
\end{align}
\begin{align}
\sigma_{32}&=\frac{\alpha(\alpha-\beta)|\Omega|^2+4 i ( \gamma_{21}+i \delta)(w_{43}-i \gamma_{42}+\delta+\Delta)}{Z}\Omega_{32},\label{sigma32}
\end{align}
\noindent where the denominator $Z$ takes the form
\begin{align}
Z&=8(\gamma_{21}+i \delta)(\omega_{43}-i \gamma_{42}+\delta+\Delta)[\gamma_{32}+i(\Delta+\delta)]\nonumber\\
      &\quad+2|\Omega|^2[\omega_{43}-i\gamma_{42}+\delta+\Delta+\alpha^2 (-i \gamma_{32}+\delta+\Delta)],
\end{align}
\noindent and $|\Omega|^2=\Omega_{31}\Omega_{13}$. Finally, we calculate the susceptibilities for the $\ket{2}$ to $\ket{3}$ and $\ket{2}$ to $\ket{4}$ transitions from Eqs.~(\ref{sigma42}) and~(\ref{sigma32}) through the relation
\begin{align}
\chi_{ij}=\frac{ 2 \mathcal{N} d_{ij}}{\epsilon_{0} \hbar \Omega_{ij} }\sigma _{ij},
\end{align}
where $\mathcal{N}$ is number density of the atomic medium.

\section{\label{app:B}Doppler broadening of susceptibilities}

Following the procedure presented in~\cite{EITXiao1}, we obtain analytical expressions for the susceptibilities after taking Doppler broadening into account. For an atom with velocity $v$ that is  interacting with co-propagating control and probe fields of very close frequency, we can take the Doppler shift into account by considering that it only acts  on the one-photon detuning according to
\begin{align}
\Delta \to \frac{v\omega_{c}}{c}+\Delta,
\end{align}
where $\omega_{c}$ is the frequency of the control field, and by replacing the number density with an atomic density distribution given by
\begin{align}
\mathcal{N}(v) \to \frac{\mathcal{N}_{0}}{\sqrt{\pi } u}e^{-v^2/u^2},
\end{align}
where $u/\sqrt{2}$ is the root mean square velocity of the atomic vapor. Finally, after taking these changes into account, we integrate the susceptibility over the velocity distribution,
\begin{align}
\chi_{ij}^{\rm Dopp}=\int_{-\infty}^{\infty} \chi_{ij}(v) dv,
\end{align}
to obtain an analytical solution for the Doppler broadened susceptibilities, $\chi_{32}$ and $\chi_{42}$. After Doppler broadening, the susceptibilities take the form
\begin{widetext}
\begin{align}
\chi_{ij}^{\rm Dopp}&=A_{ij}\frac{e^{-a_2^2/u^2}(K_{ij}+a_2)~\text{Erfc}[ia_2/u]-e^{-a_1^2/u^2}(K_{ij}+a_1)~\text{Erfc}[i a_1/u]}{a_1-a_2},
\end{align}
with
\begin{align}
a_{1,2}&=\frac{i \left[\left(\alpha ^2+1\right) |\Omega|^2+4 (\gamma _{21}+i \delta ) (\gamma _{32}+\gamma _{42}+2 i (\delta +\Delta )+i  \omega_{43})\right]}{8 \omega _c/c (\gamma _{21}+i \delta )} \nonumber\\
&\qquad\pm \frac{\sqrt{-\left[(\alpha -i)^2 |\Omega|^2+4 (\gamma _{21}+i \delta ) (-\gamma _{32}+\gamma _{42}+i  \omega_{43})\right] \left[(\alpha +i)^2 |\Omega|^2+4 (\gamma _{21}+i \delta ) (-\gamma _{32}+\gamma _{42}+i \omega_{43})\right]}}{8 \omega _c/c (\gamma _{21}+i \delta )},
\end{align}
\end{widetext}
where the positive (negative) sign corresponds to $a_{1}$ ($a_{2}$), Erfc($z$) is the complementary error function, and the coefficients $A_{ij}$ and $K_{ij}$ for susceptibility $\chi_{32}$ and $\chi_{42}$ are defined as
\begin{align}
A_{42}&=\frac{c d_{42}^2 \mathcal{N}_{0}}{\sqrt{\pi } \mu  \epsilon _0 \hbar  \omega _c},\\
A_{32}&=\frac{c d_{32}^2 \mathcal{N}_{0}}{\sqrt{\pi } \mu  \epsilon _0 \hbar  \omega _c},
\end{align}
\begin{align}
K_{42}&=\!-i c \left[\frac{|\Omega ^2| \alpha  (\alpha\! -\beta )}{4 \left(\gamma _{21}\!+i \delta \right) \omega _c}\!+\frac{\gamma _{42}\!+i (\delta\! +\Delta )\!+i \omega_{43}}{\omega _c}\right],\\
K_{32}&=\!-i c \left[\frac{ |\Omega ^2| (\beta\! -\alpha )}{4 \beta  \left(\gamma _{21}\!+i \delta \right) \omega _c}\!+\frac{\gamma _{32}\!+i (\delta \!+\Delta )}{\omega _c}\right].
\end{align}

\section{\label{app:C} Dressed state picture}

To provide a physical interpretation of the results obtained from the calculated susceptibilities, we use the dressed state picture~\cite{EITRevMarangos,DressReynaud,PlasonGong}. If we take into account that the control field can couple one of the ground states to the two excites states, we can write the Hamiltonian for the subsystem composed of these three energy levels and the control field in the rotating frame as

\begin{align}
\hat{H} &=
\hbar  \left(
\begin{array}{ccc}
 -\Delta  & 0 & \Omega _{31}/2 \\
 0 & \omega _{43}-\Delta  & \Omega _{41}/2 \\
 \Omega _{31}/2 & \Omega_{41}/2 & 0 \\
\end{array}
\right).
\end{align}
We can use this Hamiltonian to obtain eigenvalues for the atomic plus control field system. We find the eigenvalues to be given by $\frac{\hbar}{2} \lambda$, where $\lambda$ is the solution to the following cubic equation
\begin{align}
\lambda ^3+\lambda ^2 \left(4 \text{$\Delta $}-2 \omega _{43}\right)+\lambda  \left(4 \text{$\Delta $}^2-4 \text{$\Delta $} \omega _{43}-\Omega _{31}^2-\Omega _{41}^2\right)\nonumber\\-2 \text{$\Delta $} \Omega _{31}^2-2 \text{$\Delta $} \Omega _{41}^2+2 \omega _{43} \Omega _{31}^2=0.
\end{align}
with  corresponding eigenvectors
\begin{align}
   \left\{\frac{\Omega _{31}}{2 \Delta +\lambda },-\frac{-2 \Delta  \lambda -\lambda ^2+\Omega _{31}^2}{\Omega _{41} (2 \Delta +\lambda )},1\right\}.
\end{align}
These solutions capture most of the essential physics of the four-level system.

\end{document}